\documentclass{aa}  
\usepackage{orcidlink}
\usepackage{graphicx}
\usepackage{txfonts}

\newcommand{\dg}{$^{\circ}$}
\newcommand{\alfven}{Alfv\'en}
\newcommand{\rsun}{$R_\odot$}

\begin{document}

   \title{Three-dimensional properties of a coronal shock and the longitudinal distribution of its related solar energetic particles}
   \titlerunning{Shock properties and the longitudinal distribution of its related solar energetic particles}

   \author{
       Yue Zhou\inst{1,2}~\orcidlink{0000-0002-3341-0845}
        \and Li Feng\inst{1,2}\thanks{Corresponding author: lfeng@pmo.ac.cn}~\orcidlink{0000-0003-4655-6939}
        \and Guanglu Shi\inst{1}~\orcidlink{0000-0001-7397-455X}
        \and Jingnan Guo\inst{3,4}~\orcidlink{0000-0002-8707-076X}
        \and Liuguan Ding\inst{5}
        \and Yi Yang\inst{6}
        \and Jianchao Xue\inst{1}~\orcidlink{0000-0003-4829-9067}
        \and Jun Chen\inst{1}~\orcidlink{0000-0003-3060-0480}
        \and Weiqun Gan\inst{1,7}~\orcidlink{0000-0001-9979-4178}
        }

   \institute{Key Laboratory of Dark Matter and Space Astronomy, Purple Mountain Observatory, Chinese Academy of Sciences, Nanjing 210023, China
   \and School of Astronomy and Space Science, University of Science and Technology of China, Hefei 230026, China
   \and National Key Laboratory of Deep Space Exploration/School of Earth and Space Sciences, University of Science and Technology of China, Hefei 230026, China
   \and CAS Center for Excellence in Comparative Planetology, University of Science and Technology of China, Hefei 230026, China
   \and School of Physics and Optoelectronic Engineering, Institute of Space Weather, Nanjing University of Information Science and Technology, Nanjing 210044, China
   \and State Key Laboratory of Solar Activity and Space Weather, National Space Science Center, Chinese Academy of Sciences, Beijing 100190, People’s Republic of China
   \and University of Chinese Academy of Sciences, Nanjing 211135, China}

   \date{Received XXX; accepted XXX}

  \abstract
   {The widespread longitudinal distribution of solar energetic particles (SEPs) is influenced by magnetic connectivity from the observers to coronal mass ejection (CME)-driven shocks. This connectivity determines shock properties encountered by magnetic-field lines, which in turn regulate the initial particle injection and acceleration efficiency.}
   {We aim to investigate the relationship between the spatial–temporal evolution of shock properties and the longitudinal dependence of SEP intensities and spectra.} 
   {The shock parameters, including the normal speed, oblique angles, compression ratio, and \alfven\ Mach number, were derived by combining a steady-state solar-wind simulation with the three-dimensional (3D) reconstruction of the shock surface based on multi-view observations. We compared the local shock parameters at the magnetic connecting points with in situ proton intensities and peak spectra to establish the link between shock evolution and SEP characteristics.}
   {The shock nose consistently exhibited higher particle-acceleration efficiency with the largest normal speed, compression ratio, and supercritical {\alfven} Mach number, while the flanks showed delayed transition to supercritical {\alfven} Mach number with weaker efficiency. The earliest and most rapid proton enhancement of STEREO-B correlated with efficient shock acceleration and prompt magnetic connectivity to the shock. Spectral analysis revealed that proton energy spectra were consistent with the relativistic diffusive shock acceleration (DSA) estimations.}
   {The initial shock acceleration began at about $1.4 \sim 5$ {\rsun} and caused the widespread longitudinal SEP distribution. The longitudinal dependence of SEP intensity and spectral variations arise from the combined influence of 3D shock properties, magnetic connectivity, and particle transport processes. The agreement between in situ proton indices and relativistic DSA estimations supports DSA in this SEP event and provides insights into the early-stage acceleration at the source region.}

   \keywords{Sun: coronal mass ejections (CMEs) --
                shock waves --
                Sun: particle emission --
                Sun: heliosphere
               }
   \maketitle

\nolinenumbers
\section{Introduction}
Shocks driven by coronal mass ejections (CMEs) can accelerate particles from several kiloelectronvolts up to several gigaelectronvolts, and produce gradual SEP events \citep{Reames1995, Zank2000}. Compared to impulsive SEP events originating from magnetic reconnection in solar flares or jets, gradual SEP events are characterized by longer durations and broader longitudinal distributions \citep{Reames1999}. The primary acceleration mechanism for gradual SEPs is DSA \citep{Lee1983, Zank2015}, wherein particles gain energy through repeated interactions with magnetic fields in the upstream and downstream regions of shock fronts. The efficiency of DSA depends on several critical shock parameters, including the normal speed, compression ratio, \alfven\ Mach number, and the angle between the upstream magnetic field and shock normal \citep[e.g.,][]{Bemporad2011, Lario2017, Kouloumvakos2019}. Simulations have demonstrated that higher shock-normal velocities, larger compression ratios, and more oblique angles can enhance acceleration efficiency \citep{Kozarev2016}. However, the particle-acceleration process is complex and may be further modulated by additional factors, such as scattering during transport or preexisting interplanetary turbulence \citep[e.g.,][]{Gopalswamy2004, Desai&Giacalone2016}. These compounding effects present significant challenges in establishing direct correlations between in situ SEP observations and their source acceleration processes.

Multi-view observations have revealed substantial variations in particle intensity profiles during individual SEP events across different longitudinal and radial locations \citep[e.g.,][]{Lario2013, Cohen2017}. This spatial dependency is primarily attributed to differences in magnetic connectivity to the CME-driven shock front and the resulting particle-acceleration efficiency \citep{Cane1988, Reames1999}. Observers magnetically connected to regions near the shock nose typically detect a sharp and rapid intensity enhancement, reflecting more efficient particle acceleration at these locations. Conversely, observers connected to the shock flanks generally record a more gradual rise in intensity due to weaker acceleration conditions. 

The three-dimensional (3D) geometry of shocks combined with magnetohydrodynamic (MHD) simulations provides critical insights into shock properties and their evolution. Based on the multi-perspective observations from the Solar Terrestrial Relations Observatory (STEREO-A and STEREO-B; hereafter STA and STB) and near-Earth spacecraft, shock surfaces are often approximated as spherical or ellipsoidal geometries \citep[e.g.,][]{Hess2014, Kwon2014}. However, these idealized geometric representations often inadequately capture the true irregular geometry of shock fronts, which vary among events and thus introduce significant uncertainties in the calculation of shock parameters. To improve this, \cite{Feng2020} developed the mask-fitting method without a priori assumption of the shock shape. The method has also been successfully applied to the 3D reconstruction of CMEs \citep{Feng2012, Feng2013, Ying2022}.

The SEP event on 2012 March 7 was observed over a broad range of heliolongitude and has been widely analyzed in previous studies. Enhanced proton and electron intensities were detected by multiple spacecraft located at different longitudinal and radial positions, including STEREO, Solar and Heliospheric Observatory (SOHO), MErcury Surface, Space ENvironment, GEochemistry, and Ranging (MESSENGER), and Mars Science Laboratory spacecraft \citep[e.g.,][]{Ding2016, Lario2013, Zeitlin2013}. \cite{Ajello2014} and \cite{Jin2024} also analyzed the long-duration high-energy gamma-ray emission of this event detected by the Fermi Large Area Telescope (LAT). As an extension of the work by \cite{Feng2020}, this study aims to investigate how the spatial and temporal evolution of shock properties influences early-phase particle acceleration across different heliolongitudes. The structure of the paper is as follows. Section \ref{sec:model} describes the steady-state background solar-wind model, which provides upstream shock parameters and magnetic connectivity. Section \ref{sec:shock} analyzes the spatio-temporal evolution of shock parameters. Section \ref{sec:lon} presents the longitudinal distributions of the SEP event, and their relationship with shock properties is discussed in Section \ref{sec:dis}, followed by the conclusion in Section \ref{sec:con}.

\begin{figure*}[htb!]
    \begin{center}
    \includegraphics[width=0.8\textwidth]{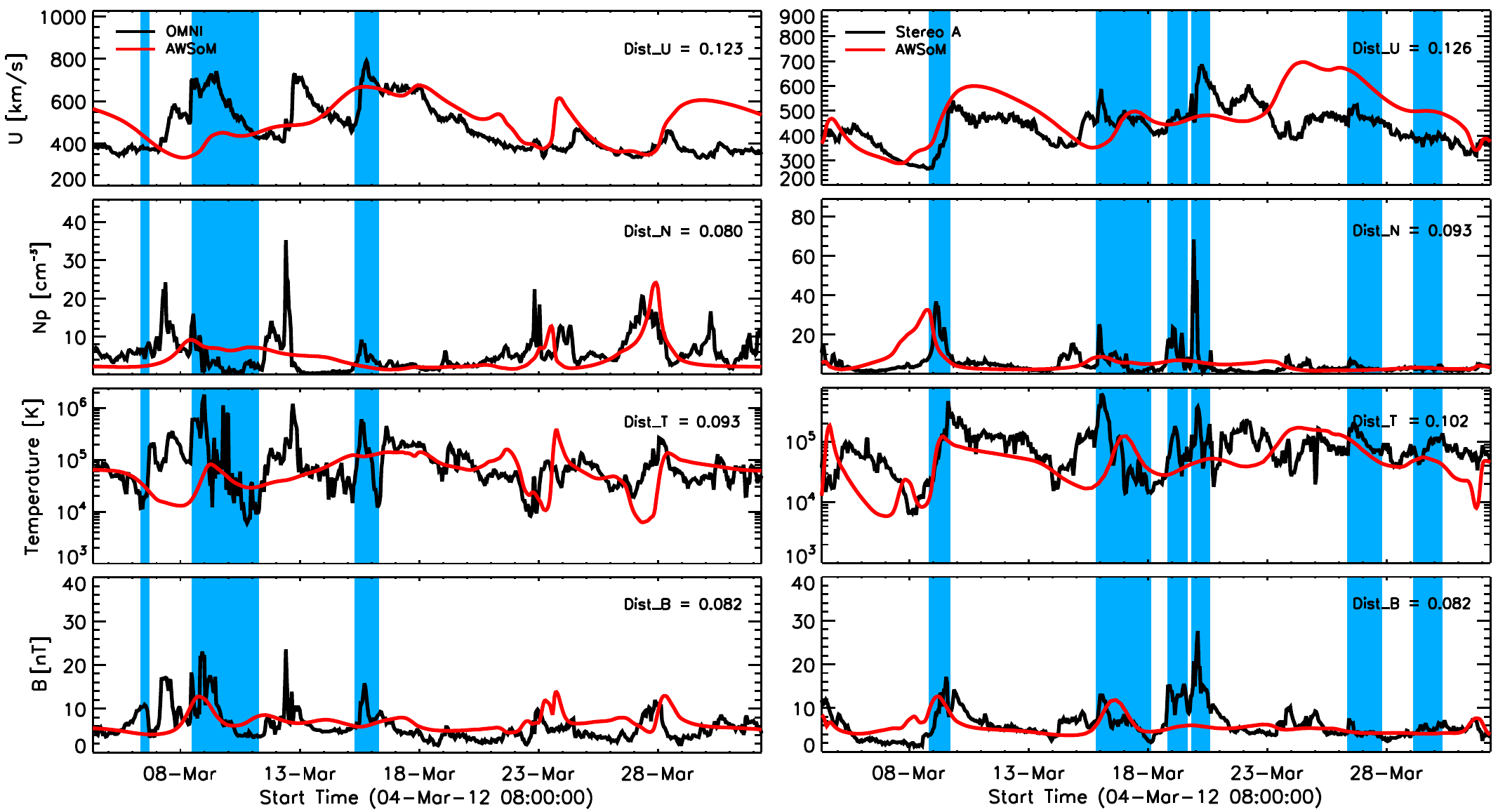}
    \caption{Comparison of solar-wind velocity, proton density, proton temperature, and magnetic fields observed by OMNI (left panel) and STA (right panel) with the steady-state solar-wind model for Carrington rotation 2121. The blue shading represents the passage of ICMEs.  \label{fig:fig1}}
    \end{center}
\end{figure*}

\section{3D reconstruction of shock surface}
We used the mask-fitting technique to reconstruct the 3D geometry of the CME-driven shock associated with the 2012 March 7 SEP event. The reconstruction is based on multi-view extreme-ultraviolet (EUV) and white-light observations. The outermost boundary of the shock surface was first manually identified in running-difference or enhanced images for each viewpoint and converted into binary masks in the corresponding image planes. To estimate the uncertainty inherent in this manual process, we repeated the tracing five times. The resulting maximum displacement along the radial direction was approximately 0.03 {\rsun}. 

We then defined a 3D Cartesian grid around the Sun in the Carrington coordinate system and projected each grid point back onto all image planes. Grid points whose projections consistently fell inside all the shock masks were considered to belong to the shock surface. After identifying all 3D points, the convex hull was applied to constrain the outer envelope of the reconstructed surface. This polyhedral boundary was smoothed using Bézier splines to produce a continuous representation of the shock surface. To compensate for projection effects and temporal offsets between multi-spacecraft observations, epipolar geometry was applied. 

Each set of 2D masks was processed through the whole mask‑fitting pipeline and produced five independent 3D shock surfaces. The maximum radial deviation among these final 3D reconstructions was approximately 0.8 {\rsun}. This approach avoids the need to adopt an assumed geometric shape and enables the reconstructed surface to capture asymmetries and local curvature variations of the shock. More details can be seen in \cite{Feng2012, Feng2020}.

\section{Background solar-wind model} \label{sec:model}
To infer the parameters of upstream shocks, we performed a data-driven 3D MHD simulation of background solar wind using the improved {\alfven} Wave Solar atmosphere Model \citep[AWSoM,][]{Holst2022}, integrated within the space weather modeling framework \citep[SWMF,][]{Toth2012}. AWSoM self-consistently describes the physical processes of coronal heating and solar-wind acceleration through {\alfven} waves based on MHD theory. The model nonlinearly couples low-frequency {\alfven} waves that propagate along and against the magnetic-field lines and results in turbulent cascades and dissipative heating \citep{Velli1989, Zank1996, Matthaeus1999, Chandran2011, Zank2012}. To separate the temperature of electrons and protons, the AWSoM model considers collision heat conduction for electron heating and radiation loss based on the Chianti model \citep{Dere1997}. Through the physical mechanism of {\alfven} wave reflection and stochastic heating, the anisotropic (parallel and perpendicular) proton temperature and the kinetic instability processes were included in the simulation \citep{Holst2014ApJ, Meng2015, Holst2019, Holst2022}.

The steady-state solar wind was simulated by the SC (1.0-24.0 {\rsun}) and IH (18.0-250.0 {\rsun}) components in the SWMF, which uses the Block-Adaptive-Tree-Solarwind-Roe-Upwind-Scheme \citep[BATS - R - US,][]{Powell1999} to solve the MHD governing equations. The initial distribution of the coronal magnetic fields was extrapolated by the potential field source surface \citep[PFSS, ][]{Altschuler1969, Schatten1969} model based on the spherical harmonics method \citep{Sachdeva2021}. We employed the seventh realization of the ADAPT-GONG map as the inner boundary condition. Prior to use, this map was processed by the air force data assimilation photospheric flux transport \citep[ADAPT, ][]{2010AIPC.1216..343A, 2015SoPh..290.1105H} model and remapped to a resolution of 360{\dg} in longitude and 180{\dg} in latitude. The initial temperature and number density of electrons and protons at the inner boundary were $T_\odot = 5 \times 10^4\ {\rm K}$ and $N_\odot = 2 \times 10^{11}\ {\rm cm^{-3}}$, respectively. Higher density values can effectively avoid the chromospheric evaporation effect without affecting the final result \citep{Lionello2009}. The ratio, $(S_A\,/\,B)_\odot$, between the Poynting flux $S_A$ and the magnetic-field strength $B$, is a free parameter of the model and has a significant influence on the simulated plasma parameters and coronal structure. By comparing the simulation results under different $(S_A\,/\,B)_\odot$ values with observations, it is found that when the value is set to $5.4 \times 10^5\ {\rm W\,m^{-2}\,T^{-1}}$, the consistency is better. The correlation length of the {\alfven} wave, $L_\perp$, is proportional to $B^{-1/2}$, and the proportionality constant $L_\perp \sqrt{B}$ is set as an adjustable parameter to $1.5 \times 10^5\ {\rm m\,\sqrt{T}}$. The stochastic heating exponent and amplitude \citep{Chandran2011} related to the energy partitioning between electrons and protons were set to 0.21 and 0.18, respectively.

The SC component used an adaptive, 3D, spherical stretched grid in the heliographic rotating (HGR) coordinate, while the IH component used a Cartesian grid in the heliographic corotating (HGC) coordinate. The grid blocks in the SC and IH domains were composed of $6 \times 8 \times 8$ and $8 \times 8 \times 8$ mesh cells, respectively. To improve the simulation resolution, the adaptive mesh refinement (AMR) library \citep{Toth2012} was applied to refine the grid in the regions of the lower corona (1.0-1.7 {\rsun}), the heliospheric current sheet location, and the Earth observer. The horizontal resolution of the grid in SC ranges from 0.35{\dg} to 2.81{\dg}, and the cell size in IH varies between 0.24 {\rsun} and 7.81 {\rsun}. The SC component used the second-order scheme for 90,000 iterations to obtain a steady-state corona and then ran the fifth-order scheme \citep{Suresh1997, Chen2016} for 120,000 iterations to further improve accuracy. Next, the SC and IH components were coupled in a butterfly grid in the range of 18.0-20.0 {\rsun} for one step to transfer the SC simulation results to IH. 15,000 iterations were then performed in IH to obtain the steady-state solar wind as well.

Figure \ref{fig:fig1} displays the steady-state model results compared to in situ solar-wind velocity, proton density, proton temperature, and magnetic fields observed by OMNI and STA. The OMNI data set \citep{King2005} compiles solar-wind and interplanetary-magnetic-field (IMF) parameters from multiple spacecraft located near Earth, including the Advanced Composition Explorer (ACE), Wind, Geotail, and IMP8. The hourly averaged plasma parameters of STA were obtained from the Space Physics Data Facility\footnote[1]{\url{https://spdf.gsfc.nasa.gov}},  as well as the OMNI data set. We adopted the Dist index to characterize the error between the model results and observations \citep{Sachdeva2019}. The Dist indices of the proton density and magnetic fields served as the evaluation criterion for the model because of their involvement in calculating shock parameters. It can be seen that the parameters of the steady-state model show good agreement with in situ observations and capture the overall trend, especially during periods unaffected by interplanetary coronal mass ejections (ICMEs), as indicated by the blue shading in Figure \ref{fig:fig1}. ICME events are identified by the conjunction of a preceding (2–3 days) halo CME propagating toward the spacecraft and the significant plasma parameter enhancements from background levels. The event duration is defined as the period from the initial rise of plasma parameters until their recovery to background levels.

\begin{figure*}[ht!]
    \begin{center}
    \includegraphics[width=0.8\textwidth]{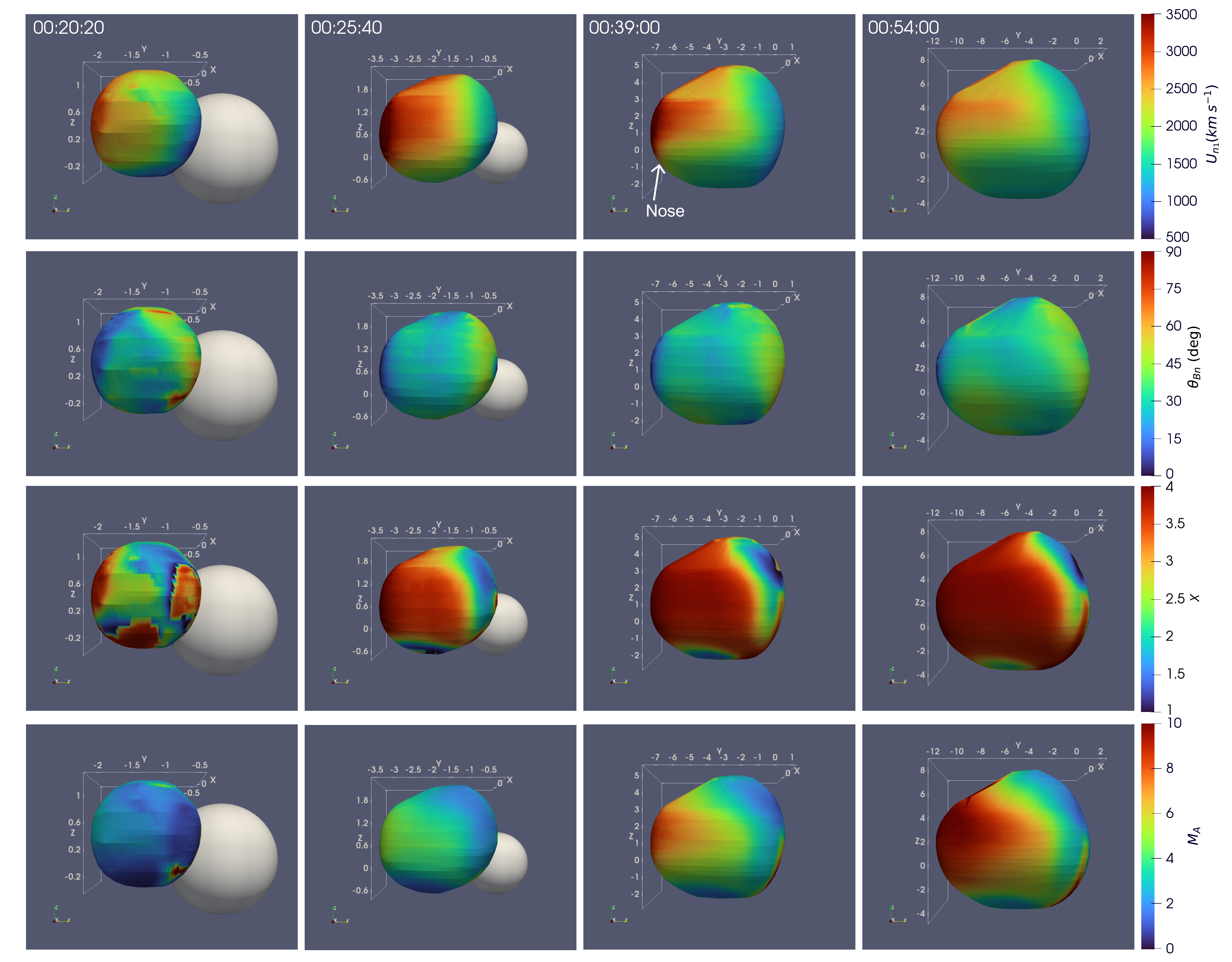}
    \caption{Spatio-temporal evolution of shock-normal speed, oblique angle, compression ratio, and \alfven\ Mach number from top to bottom, respectively (see Movie 1 for the dynamic evolution). The gray sphere represents the Sun. Coordinates are defined in the HGC frame, with the negative X-axis directed toward Earth at 18:00 UT on 2012 March 18. The axes are in units of the solar radius. \label{fig:fig2}}
    \end{center}
\end{figure*}

\begin{figure*}[ht!]
    \begin{center}
    \includegraphics[width=0.8\textwidth]{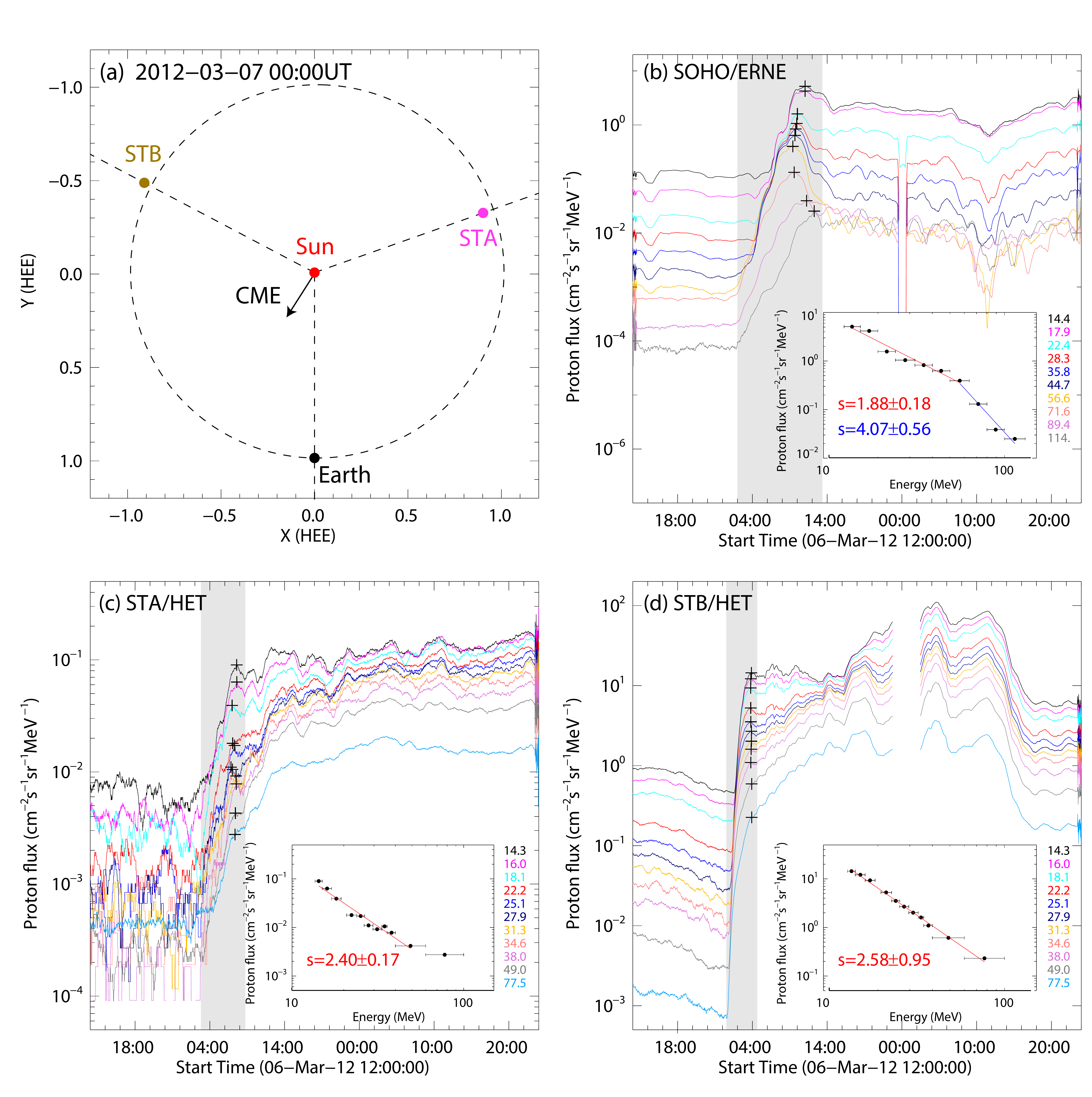}
    \caption{(a) Positions of STA, STB, and Earth at 00:00 UT on 2012 March 7. The black arrow represents the direction of the CME propagation. (b) 1-hour averaged proton intensities measured by SOHO/ERNE. (c) 1-hour averaged proton intensities measured by STA/HET. (d) 1-hour averaged proton intensities measured by STB/HET. The black plus signs represent the peak flux of each energy range. The inset panels display the power-law fitting of peak flux, and the fit spectral indices are indicated. The shaded region is the time interval used for the fluence spectrum. 
        \label{fig:fig3}}
    \end{center}
\end{figure*}

\section{Shock properties} \label{sec:shock}
The normal to the shock surface ($\mathbf{n}$) and shock-normal speed ($U_{n1}$) can be obtained from the 3D reconstructed shock point clouds. We applied the K-nearest-neighbor (KNN) technique to select 30 nearest points distributed around a given point and principal component analysis (PCA) to calculate the normal to the shock surface. The eigenvector corresponding to the smallest eigenvalue represents the normal direction. For each shock point at the time, $t_{i}$, its normal direction was further used to determine its intersection with the shock surface at the subsequent time $t_{i+1}$. According to these intersection points, we estimated the propagating distance between $t_{i}$ and $t_{i+1}$ and therefore derived the shock speed along the normal direction. The steady-state background solar-wind model provided the magnetic-field ($\mathbf{B}$), plasma-density ($\rho$), and plasma-pressure ($p$) parameters. Based on these parameters, we calculated the angle between the upstream magnetic fields and shock normal ($\theta_{Bn}$), shock \alfven\ Mach number ($M_{A}$), and compression ratio \citep[X,][]{Gurnett2017} as

\begin{equation}
        \theta_{Bn}=cos^{-1} \bigg( \frac{\mathbf{B} \cdot \mathbf{n}}{\lvert \mathbf{B} \rvert} \bigg) \label{eq1} 
\end{equation}
\begin{equation}
   M_{A}=\frac{U_{n1}}{V_{A1}cos\theta_{Bn}} \label{eq2} 
\end{equation}

\begin{align}
    (&U_{n1}^{2}-XV_{A1}^{2}cos^{2}\theta_{Bn})^{2} \left[U_{n1}^{2} \right.  \nonumber \\
    & \left. - \frac{2XV_{S1}^{2}}{X+1-\gamma(X-1)} \right] -Xsin^{2}\theta_{Bn}U_{n1}^{2}V_{A1}^{2} \cdot \nonumber \\
    & \left[\frac{2X-\gamma(X-1)}{X+1-\gamma(X-1)}U_{n1}^{2}-XV_{A1}^{2}cos^{2}\theta_{Bn} \right] =0, \label{eq3} 
\end{align}
where $V_{A1}=B/\sqrt{\mu_{0}\rho}$ and $V_{S1}=\sqrt{\gamma p/\rho}$ are the upstream \alfven\ speed and sound speed, respectively, and $\gamma=5/3$ is the adiabatic index. The spatio-temporal evolution of the shock properties is displayed in Figure \ref{fig:fig2}, with color-scales from top to bottom representing the shock-normal speed, oblique angle, compression ratio, and \alfven\ Mach number. It can be seen that the shock front expanded across the solar hemisphere and eventually encompassed almost the whole longitude. The main findings of these shock properties are listed below:

\begin{enumerate}
    \item The shock-normal speed exhibits significant spatial variation. The nose region reaches velocities of up to about 3500 $\mathrm{km\ s^{-1}}$ at approximately 2.5 {\rsun}, consistently exceeding those at the flanks. This spatial asymmetry persists throughout the shock evolution and favors sustained and efficient particle acceleration at the nose relative to the flanks. Additionally, the flank regions adjacent to the shock nose undergo a distinct deceleration after 00:39 UT. 
    \item The $\theta_{Bn}$ distribution reveals that the majority of the shock regions remain quasi-parallel over time, suggesting DSA during the early phase of this SEP event. The shock nose exhibits the smallest $\theta_{Bn}$ values, while the flanks show progressively larger angles (up to 65\dg) as the shock normal becomes more oblique to the IMF. In accordance with DSA theory, the acceleration efficiency increases with $\theta_{Bn}$ following a $sec^2(\theta_{Bn})$ dependence \citep{Kozarev2015}. Thus, the higher $\theta_{Bn}$ values correspond to more efficient acceleration, for particles already participating in diffusive acceleration. However, quasi-perpendicular shocks commonly exhibit reduced injection efficiency for thermal particles because such particles are magnetically tied to field lines and less likely to re-cross the shock without sufficient perpendicular scattering \citep{Giacalone2005, Caprioli2014}.
    \item High compression ratios ($>3$) first appear only at the shock nose and partial flanks, and then they spread throughout most shock regions as the shock evolves. The compression ratios peak at the nose and decrease gradually toward the flanks, which is consistent with the results of \cite{Bemporad2011}. Notably, localized compression-ratio enhancements emerge at specific flank regions, likely due to their larger $\theta_{Bn}$ and lower $M_{A}$.
    \item The shock nose exhibits higher $M_{A}$ values than the flanks, indicating greater shock strength at the nose. This spatial distribution correlates strongly with the faster normal speed and smaller $\theta_{Bn}$ at the shock nose. At low plasma $\beta$, the critical Mach numbers for quasi-parallel and quasi-perpendicular shocks are 1.53 and 2.76, respectively \citep{Treumann2009}. At 00:20:20 UT, while the quasi-parallel nose at $\sim 2$ {\rsun} already exceeds the critical threshold, the higher $\theta_{Bn}$ flanks remain subcritical. By 00:25:40 UT, most shock regions become supercritical, except the upper flank region. According to the simulation by \cite{Giacalone1997}, the particle-acceleration efficiency exhibits a non-monotonic dependence on $M_{A}$, peaking when $M_{A}$ approaches 10 and decreasing at higher values. In our event, $M_{A}$ remains predominantly below 10 throughout the shock front, suggesting favorable conditions for efficient particle acceleration.
\end{enumerate}

\begin{figure*}[ht!]
    \begin{center}
    \includegraphics[width=0.8\textwidth]{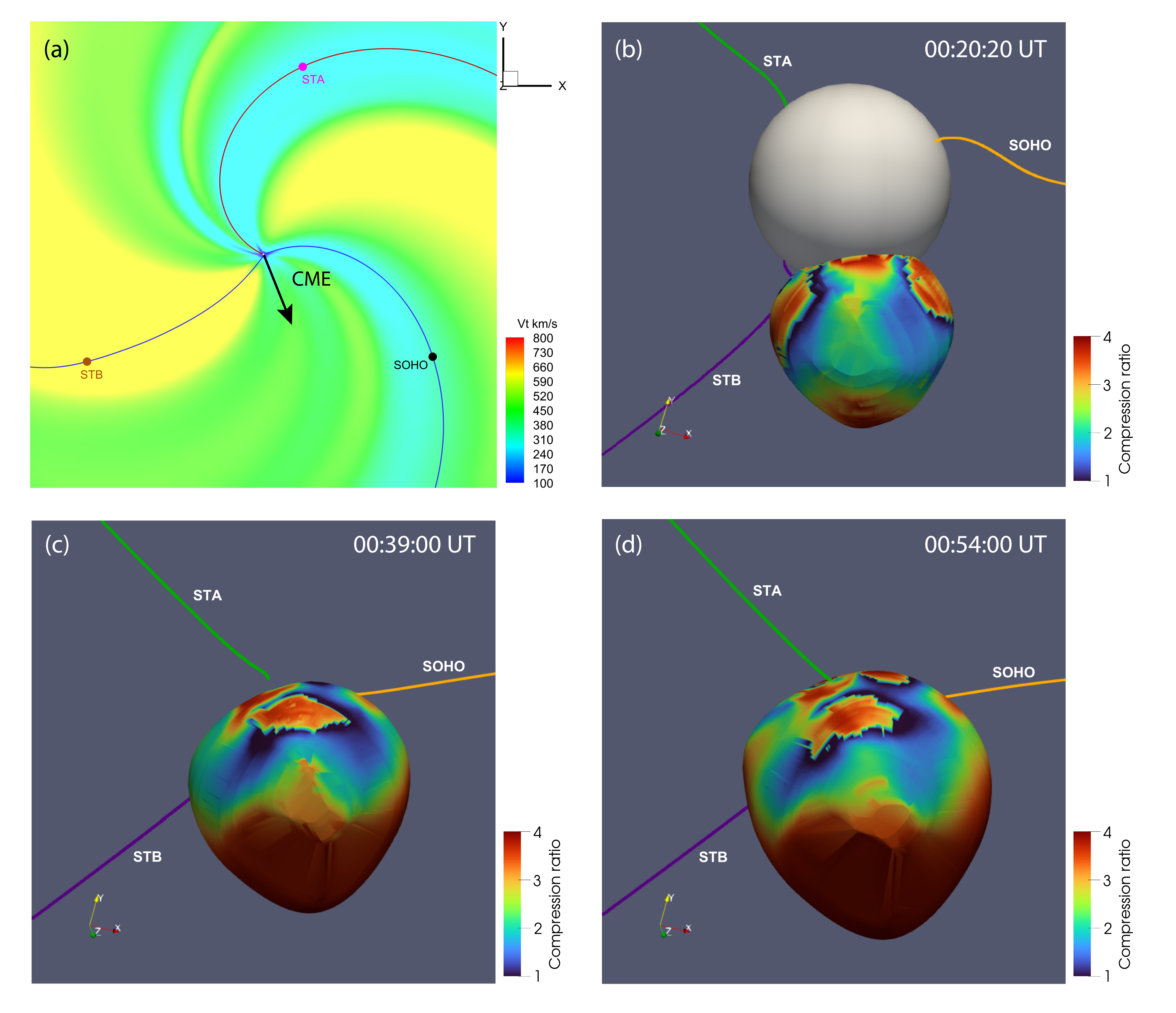}
    \caption{(a) Polar velocity in ecliptic plane from the SC and IH components during Carrington rotation 2121. The positions of STA, STB, and SOHO on 2012 March 7 are marked by pink, brown, and black dots, respectively. The black arrow represents the propagating direction of the associated CME. The solid lines represent the magnetic-field lines connecting each observer with the Sun. (b)-(d) Compression ratios of the reconstructed shock at 00:20:20 UT, 00:39:00 UT, and 00:54:00 UT are shown. The solid yellow, green, and purple lines are the magnetic-field lines connecting SOHO, STA, and STB to the SC component, respectively. The sphere represents the Sun. The coordinate system and axis scales are the same as those in Figure \ref{fig:fig2}. \label{fig:fig4}}
    \end{center}
\end{figure*}

\begin{table*}[ht!]
\caption{Properties of shock and SEPs observed at different heliolongitudes}              
\label{tab:index}    
\centering      
\renewcommand{\arraystretch}{1.8}
\begin{tabular}{c c c c c c c c c }     
\hline\hline              
 Observers & Peak index\tablefootmark{a} & Fluence index\tablefootmark{a} & $L(AU)$\tablefootmark{b} & $X$\tablefootmark{c} &  Nonrelativistic index\tablefootmark{d} & Relativistic index\tablefootmark{d}\\  
\hline                     
    SOHO  & 1.88$\pm$0.18/4.07$\pm$0.56 & 1.91$\pm$0.2/4.39$\pm$0.24 & 1.11 & 3.15$\pm$0.2 & $1.2^{+0.07}_{-0.06}$ & $2.4^{+0.14}_{-0.12}$ \\
    STB  & 2.58$\pm$0.95 & 2.52$\pm$0.05 & 1.05 & 3.55$\pm$0.09 & $1.09^{+0.02}_{-0.02}$ & $2.18^{+0.04}_{-0.04}$ \\
    STA & 2.4$\pm$0.17 & 2.44$\pm$0.11 & 1.26 & 2.91$\pm$0.77 & $1.29^{+0.53}_{-0.23}$ & $2.58^{+1.06}_{-0.46}$ \\
\hline                               
\end{tabular}
\tablefoot{\tablefoottext{a}{Spectral indices fitted using peak (plus signs in Figure \ref{fig:fig3}) and time-integrated (shaded region in Figure \ref{fig:fig3}) flux.}
\tablefoottext{b}{Path length of magnetic-field lines connecting the Sun to each observer.}
\tablefoottext{c}{Compression ratios extracted from Figure \ref{fig:fig5} at 00:54 UT.}
\tablefoottext{d}{Spectral indices estimated by shock compression ratios under nonrelativistic and relativistic DSA.}}
\end{table*}

\section{Longitudinal distribution of SEPs} \label{sec:lon}
The SEP event on 2012 March 7 was one of the most intense events of the year and was observed by multiple spacecraft spanning a wide range of heliolongitude. Figure \ref{fig:fig3}(a) displays the locations of Earth, STA, and STB. It can be seen that the heliocentric longitudinal separations of STA and STB to Earth were 109.3{\dg}W and 117.6{\dg}E, respectively, and the direction of the shock nose was separated by approximately 37.3{\dg}E in longitude from Earth. Proton intensities were measured by the Energetic and Relativistic Nuclei and Electron \citep[ERNE,][]{Torsti1995} instrument onboard near-Earth SOHO at 13-130 MeV and by the High Energy Telescope \citep[HET,][]{von2008} onboard STEREO, with an energy range of 13.6-100 MeV. 

Figure \ref{fig:fig3} displays the proton flux for the 2012 March 7 SEP event: (b) 1-hour averaged proton intensities observed by SOHO/ERNE, (c) 1-hour averaged proton intensities measured by STA/HET, and (d) 1-hour averaged proton intensities measured by STB/HET. Although all three spacecraft observed significant proton intensity enhancements, their temporal intensity profiles differed markedly. STB, located east of the shock flank, recorded a rapid intensity rise at first, followed by western SOHO and STA positioned at the back of the shock. Notably, SOHO and STA exhibited similar gradual increase profiles despite their considerable longitudinal separation. These observations imply that the shape of intensity profiles is influenced not only by longitudinal locations but also by acceleration and transport processes from the source region to the observers.

To connect the in situ observation with particle-acceleration processes of the source region, we performed a proton spectral analysis using the early peak flux at each energy channel. The peak flux spectrum is less susceptible to transport modifications, especially for energies above about 10 MeV \citep{Hollebeke1975, Wang2024}. The inset panels of Figure \ref{fig:fig3} present the proton energy spectra. They are fit by power laws of $J=J_{0}E^{-s}$, where $J_{0}$ is the normalization constant, $E$ is the proton energy, and $s$ is the spectral index. The power-law fits, shown as solid lines, are summarized in Column 2 of Table \ref{tab:index}. The proton energy spectra observed by STA and STB display single-power-law distributions with spectral indices of approximately 2.4 and 2.58, respectively. In contrast, the SOHO proton spectrum exhibits a double-power-law profile, with a spectral break at around 57 MeV. Above the spectral break, the peak flux of SOHO arrives later than that at lower energies. Particles require more time to be accelerated to higher energies, and this may lead to the delayed arrival of high-energy peak fluxes \citep{Li2025, 2025ApJ...994..242C}. However, the origin of double-power-law spectra remains debated due to the energetic particles coming from different source regions \citep{Kong2019, Yu2022} or transport effects \citep{Zhao2016, Zhao2017}.

\begin{figure*}[ht!]
    \begin{center}
    \includegraphics[width=0.8\textwidth]{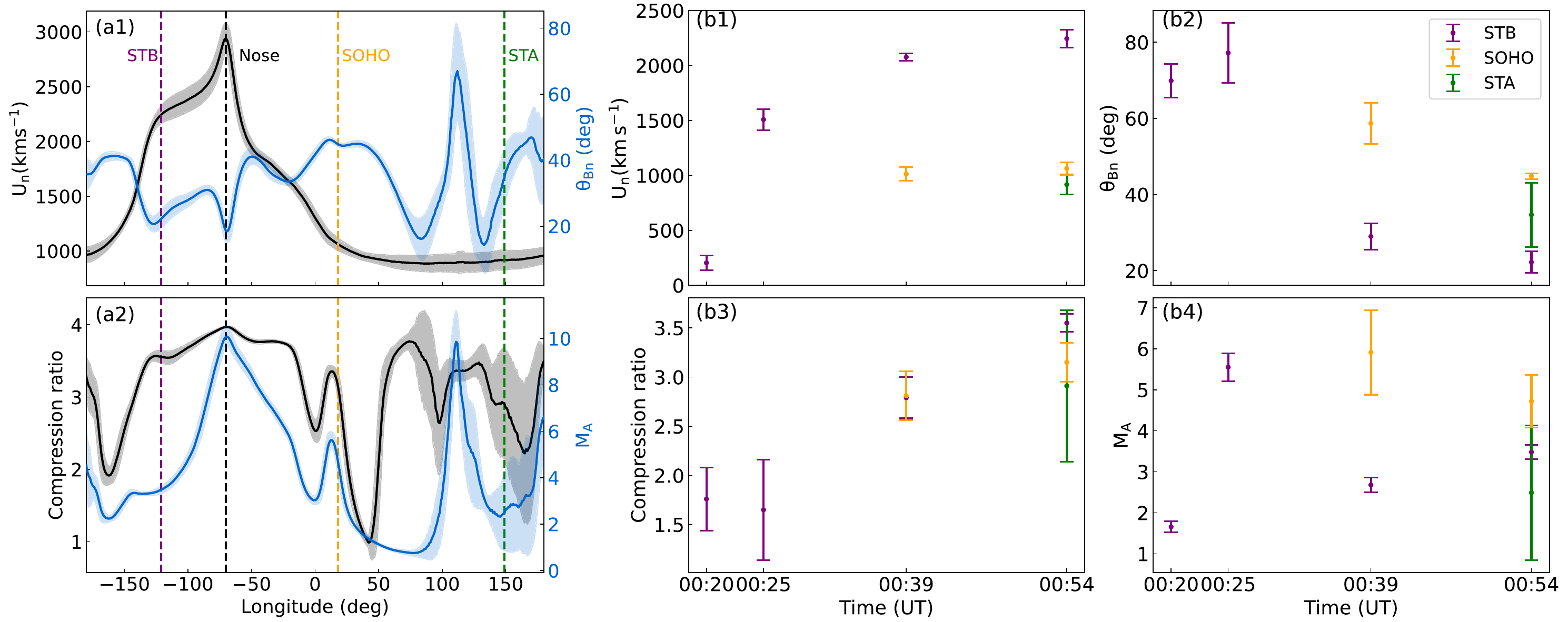}
    \caption{(a) Longitudinal distribution of shock parameters on ecliptic plane at 00:54 UT. Vertical dashed lines indicate the longitudinal locations of each observer. (b) Temporal evolution of shock parameters at 00:20:20 UT, 00:25:40 UT, 00:39 UT, and 00:54 UT. \label{fig:fig5}}
    \end{center}
\end{figure*}

\section{Discussion} \label{sec:dis}
\subsection{Magnetic connectivity and longitudinal distribution of shock properties} \label{subsec:mag}
Based on the steady-state background solar-wind model, we established the magnetic connectivity of each spacecraft (SOHO, STA, and STB) to the Sun. Magnetic-field lines were traced in the ecliptic plane from the IH and SC components and acted as Parker spiral configurations (Figure \ref{fig:fig4}(a)). SOHO, STA, and STB are magnetically connected to the western flank, shock back, and eastern flank, respectively. Therefore, we can calculate the proton propagation-path length ($L$) along the magnetic-field lines from the Sun to each observer. The length is listed in Column 4 of Table \ref{tab:index}, and it can be seen that STB has the shortest path length, followed by SOHO and STA. To account for the 3D effects in shock magnetic-field interactions, we performed additional 3D-field-line tracing. Each spacecraft's connection was first traced outward to the IH-SC coupled boundary at 20 {\rsun} in the ecliptic plane, and then sunward through the SC domain in 3D coordinates. The resulting time-dependent magnetic connectivity between each observer and the evolving shock surface is shown in Figures \ref{fig:fig4}{(b)-(d)}.

The magnetic-field line connected to STB intersects the shock front as early as 00:20:20 UT when the shock reaches approximately 1.41 {\rsun} (Figure \ref{fig:fig4}{(b)}). The expanding shock subsequently envelops the magnetic footpoint of SOHO at 00:39 UT at a greater height of about 2.74 {\rsun} (Figure \ref{fig:fig4}{(c)}), and it finally intersects the magnetic-field line of STA at 00:54 UT beyond about 1.99 {\rsun} (Figure \ref{fig:fig4}{(d)}). The precise timing of the initial magnetic connectivity is constrained by the temporal cadence of the multi-perspective white-light coronagraph observations, which allowed shock reconstruction at only four moments. Therefore, the actual initial connection may precede the first reconstructed moment at which the connection is visibly established. However, the analysis establishes a well-constrained temporal sequence of the shock evolution across regions magnetically connected to each observer. The shock first intersects the STB-connected field line at lower altitudes before reaching SOHO and STA at greater heliocentric distances.

The point magnetically connected to each observer that intersects the shock surface is the ``cobpoint'' \citep[Connecting with observer point,][]{Heras1995}. The cobpoint is the presumed site where particles are accelerated and injected into the IMF and begin to transport toward the observers. At 00:54 UT, the cobpoints of STB and SOHO are located at the shock flanks, while STA is connected to the back of the shock. Through the cobpoints, we can extract shock properties directly associated with the observer-connected regions and link them with in situ SEP observations. It should be noted that the backside geometry of the shock with respect to its propagation direction is subject to larger uncertainty from the smoothed curve inscribed in the constraining hexagon \citep{Feng2012, Feng2020}. This implies that the shock parameters of STA have great uncertainty and are thus less reliable.

Given the reconstruction uncertainties, the shock parameters were determined by first performing a latitudinal average within $\pm$0.5 {\rsun} of the ecliptic plane for each shock surface and then averaging across the five surfaces. The uncertainty incorporates the standard deviation from the latitudinal averaging and the scatter across the five reconstructed shock surfaces. The asymmetric expansion of the shock front allowed the full 360$^\circ$ longitudinal profile of shock parameters to be obtained at 00:54 UT, as shown in Figure \ref{fig:fig5}{a}. The delayed magnetic connectivity of SOHO and STA also meant that shock parameters at all cobpoints could only be derived at this moment. The shock nose is characterized by the highest normal speed, compression ratio, and {\alfven} Mach number, along with the lowest $\theta_{Bn}$. These combined factors suggest favorable conditions for efficient particle acceleration. Although higher $\theta_{Bn}$ generally facilitates particle acceleration to higher energies, quasi-parallel shocks with high speed and compression ratio can enhance low-energy fluxes and lead to harder low-energy spectra \citep{Kozarev2016}. In practice, particle acceleration further depends on local plasma conditions such as the level of upstream turbulence, foreshock region, and the availability of a seed particle population.

The shock parameters at each cobpoint were extracted from the longitudinal profile, and their temporal evolution is presented in Figure\ref{fig:fig5}{b}. STB exhibits the most favorable acceleration conditions, with the highest shock-normal speed and compression ratio. While SOHO exhibits a slightly lower compression ratio, its combination of larger $\theta_{Bn}$ and higher $M_{A}$ enables effective acceleration under the DSA theory. The confluence of these shock properties makes it challenging to compare acceleration efficiency between SOHO and STB definitively.

\subsection{Relation to shock properties} \label{subsec:relation}
The shock properties observed between 00:20 UT and 00:54 UT primarily characterize the early-stage particle-acceleration process. Although the temporal evolution of proton intensities depends on shock properties, it is also modulated by magnetic connectivity and interplanetary transport conditions. STB displays a more rapid intensity increase compared to the gradual increase observed at SOHO, despite both spacecraft being magnetically connected to shock flanks. The rapid enhancement at STB typically suggests prompt and efficient particle acceleration and injection into the IMF. However, as discussed in Section \ref{subsec:mag}, the shock properties at the cobpoints of STB and SOHO show comparable acceleration efficiency. This implies that local shock properties and acceleration conditions alone cannot explain the observed intensity differences. Additional processes, such as particle injection and transport, may play an important role in shaping the temporal profiles.

The Solar Terrestrial Relations Observatory B (STB) was magnetically connected to the shock surface as early as 00:20:20 UT, approximately 20 and 40 minutes earlier than SOHO and STA, respectively. During this initial phase, the shock parameters at the cobpoint of STB enabled the efficient and sustained particle acceleration. By 00:54 UT, $\theta_{Bn}$ at STB's cobpoint decreased to 22.23{\dg}, and this quasi-parallel shock further facilitated particle injection along the IMF. The transition from quasi-perpendicular to quasi-parallel shocks was also observed in \cite{Jin2018} due to closed magnetic fields. Furthermore, as shown in Column 4 of Table \ref{tab:index}, the shortest propagation path length of STB reduced particle transport time and contributed to its earlier onset of intensity enhancement. 

The Solar Terrestrial Relations Observatory A (STA) established the magnetic connection with the shock front at 00:54 UT, and its cobpoint conditions remained unfavorable for efficient particle acceleration, especially with the subcritical $M_{A}$. While potential uncertainties in our shock reconstruction may affect the precise shock parameters at STA, the nearly simultaneous flux increases observed at STA and SOHO indicate the role of additional acceleration mechanisms. The deduction is further supported by the weak anisotropy measured at STA, as reported by \cite{Dresing2014}. Particles were not injected locally but arrived through cross-field diffusion \citep{Kouloumvakos2016}. Besides, the strong anisotropy at STB and SOHO indicates their direct connection to shock acceleration, which is consistent with the above discussion of their favorable cobpoint shock properties.

Standard DSA theory demonstrates that the spectral index of accelerated particles is determined by the compression ratio, with distinct formulae for nonrelativistic ($s_{nr} = (X + 2)/(2X - 2)$) and relativistic ($s_{r} = (X + 2)/(X - 1)$) particles \citep{Forman1985}. Columns 6 and 7 of Table \ref{tab:index} show the theoretical spectral indices at all three observers calculated from compression ratios for nonrelativistic and relativistic regimes, respectively. It can be seen that the peak spectral indices show better agreement with the relativistic DSA predictions. Furthermore, we compared the fluence spectra integrated from the onset through the peak. The fluence spectral indices are listed in Column 3 of Table \ref{tab:index} and are similarly consistent with relativistic DSA prediction. This consistency reflects that relativistic effects become non-negligible for particle spectra in the energy range of several tens of megaelectronvolts. Protons are no longer strictly nonrelativistic at the lowest energy bin ($\sim13$ MeV) and begin to exhibit near-relativistic behavior, especially at higher energies. The transition to near-relativistic kinematics explains the observed spectral evolution and validates the use of relativistic DSA theory for interpreting high-energy particle spectra in SEP events. 

Although the spectral index of STA is closer to the relativistic DSA prediction, this agreement does not imply that the particles observed at STA were accelerated via the DSA mechanism. The shock parameters at STA are less reliable, as discussed in Section \ref{subsec:mag}. Besides, the cross-diffusion effect may attenuate contributions from other efficient acceleration regions with different spectral properties and thus result in a single-power-law spectrum at STA. In contrast, the double-power-law spectrum observed at SOHO indicates that its particle population may originate from multiple acceleration sources or has undergone modification during transport. Therefore, applying DSA theory based on compression ratios from a single connection point to such a composite spectrum leads to a larger deviation. It reflects that DSA-based spectral predictions are most reliable when applied to spectra dominated by a single acceleration source, such as STB in this event.

The deviations from idealized relativistic DSA theory also imply the presence of additional modulating factors, including (1) transport effects during interplanetary propagation from the low or mid corona to 1 AU; (2) the standard DSA theory neglects the particle scattering and $\theta_{Bn}$ \citep{Reames2012} while compression ratios calculated in this study are influenced by $\theta_{Bn}$; (3) protons may be accelerated at different locations of the shock front since magnetic connectivity was derived from the background solar-wind model rather than the true space environment under dynamic CME-driven conditions; (4) the source-region spectra, derived using the proxy of peak spectra, neglect the particle transport and scatter effects. Taken together, these results reflect the complexity of interpreting SEP spectra and temporal profiles and emphasize the role of shock properties, magnetic connectivity, and transport effects in particle acceleration.

\section{Conclusions} \label{sec:con}
Multi-spacecraft observations from SOHO, STA, and STB, with a longitudinal separation of approximately 110{\dg}, enabled the reconstruction of the 3D CME-driven shock and promoted the investigation of the longitudinal dependence of the SEP event on 2012 March 7. The shock front had expanded to about 13.5 {\rsun} by 00:54 UT, covering the low to middle corona and extending across a wide range of heliolongitude. This broad coverage allowed us to assess early-phase shock acceleration based on shock properties, including the shock-normal speed, the angle between the upstream magnetic fields and shock normal, the compression ratio, and the {\alfven} Mach number.

Throughout the shock evolution, the shock nose exhibited the highest acceleration efficiency, which is characterized by the largest normal speed, compression ratio, and {\alfven} Mach number. The shock nose was already supercritical as early as 00:20:20 UT at approximately 2.5 {\rsun}. In contrast, the flanks became supercritical roughly 20 minutes later at a greater heliocentric distance of about 3.3 {\rsun}, indicating delayed particle acceleration in these regions. This difference highlights the spatio-temporal asymmetry in shock evolution and suggests that particle acceleration can begin at different heights depending on shock geometry and properties. The shock front intersected the cobpoint of STB as early as 00:20:20 UT at about 1.4 {\rsun} and evolved until about 5 {\rsun} by 00:39 UT before the proton release time \citep[$\sim$ 00:48 UT,][]{Kouloumvakos2016}. This suggests that effective acceleration likely occurred within these heliocentric distances. We therefore conclude SEP acceleration probably began at 2.5 {\rsun} near the shock nose and between $1.4 \sim 5$ {\rsun} along the flanks. This range is in good agreement with the $2 \sim 4$ {\rsun} acceleration heights obtained from a velocity dispersion analysis \citep{Reames2009} and with the statistical range of $1.7 \sim 4$ {\rsun} derived from direct shock observations \citep{2012SSRv..171...23G} for ground level enhancement events.

Despite all three spacecraft detecting enhanced SEP intensities, their temporal profiles varied significantly. STB showed the most rapid intensity increase with prior magnetic connectivity and a shorter particle-propagation path compared with the gradual intensity rises of SOHO and STA. The results demonstrate that SEP longitudinal distributions depend not only on the cobpoint location and acceleration efficiency decided by magnetic connectivity \citep{Cane1988, Lario2013}, but also on particle-injection timing, and transport effects. A spectral analysis reveals that all three proton spectra in the energy range between about 10 and 100 MeV match relativistic DSA predictions. This agreement provides crucial insights into early-stage shock-acceleration processes in the absence of direct coronal measurements. The compression ratio at the source region can be inferred from the observed particle spectra.

In this study, magnetic connectivity was established using the background solar-wind model, which does not account for dynamic solar eruptions or interplanetary turbulence. It may result in different cobpoint locations from the present. Future work will incorporate MHD simulations of CME-driven shocks coupled with particle transport models \citep[e.g.,][]{Jin2018, Li2021, Zhao2024}. This will improve the accuracy of shock properties and magnetic connectivity and provide deeper insight into SEP acceleration and transport processes.

\begin{acknowledgements}
We sincerely thank the anonymous referee for the valuable comments that helped improve the manuscript. We thank Fang Shen for helpful discussions on the solar-wind simulations, and thank the open data policy of STEREO and SOHO. This work is supported by the Strategic Priority Research Program of the Chinese Academy of Sciences (Grant No. XDB0560000), National Key R$\&$D Program of China (Grant No. 2022YFF0503003), China's Space Origins Exploration Program(GJ11020215, GJ11020408, GJ11020405), and the National Natural Science Foundation of China (NSFC; Grant No. 12233012, 12203102, 42274215, 42188101, 42521007, 42474221).
\end{acknowledgements}

\bibliographystyle{aa} 
\bibliography{ref}

\end{document}